# The Isgur-Wise Function from the Lattice

S.P. Booth, K.C. Bowler, N.M. Hazel, D.S. Henty, H. Hoeber, R.D. Kenway,

D.G. Richards, H.P. Shanahan, J.N. Simone, A.D. Simpson

*Physics Department, The University of Edinburgh, Edinburgh EH9 3JZ, Scotland*

L. Lellouch, J. Nieves, C.T. Sachrajda, H. Wittig

*Physics Department, The University, Southampton SO9 5NH, UK*

(UKQCD Collaboration)

(September 6, 1994)

## Abstract

We calculate the Isgur-Wise function by measuring the elastic scattering amplitude of a $D$ meson in the quenched approximation on a $24^3 \times 48$ lattice at $\beta = 6.2$, using an $O(a)$-improved fermion action. Fitting the resulting chirally-extrapolated Isgur-Wise function to Stech's relativistic-oscillator parametrization, we obtain a slope parameter $\rho^2 = 1.2^{+7}_{-3}$. We then use this result, in conjunction with heavy-quark symmetry, to extract $V_{cb}$ from the experimentally measured $\bar{B} \to D^* l \bar{\nu}$ differential decay width. We find $|V_{cb}|\sqrt{\tau_B/1.48\text{ps}} = 0.038^{+2+8}_{-2-3}$, where the first set of errors is due to experimental uncertainties, while the second is due to the uncertainty in our lattice determination of $\rho^2$.

Typeset using REVTEX

Heavy-quark symmetry enables all the non-perturbative, strong-interaction physics for semi-leptonic $B \to D$ and $D^*$ decays to be parametrized in terms of a single universal function of $\omega \equiv v \cdot v'$, where $v$ and $v'$ are the four-velocities of the $B$ and $D$ mesons respectively [1,2]. This function, $\xi(\omega)$, known as the Isgur-Wise function, is normalized at the zero-recoil point: $\xi(1) = 1$ [2]. In order to determine the element $V_{cb}$ of the Cabibbo-Kobayashi-Maskawa (CKM) matrix from experimental measurements of these semi-leptonic decays (which are made at $\omega > 1$), it is necessary to know the Isgur-Wise function, and in particular its slope at the zero recoil point, $\xi'(1)$. In this letter we report on a lattice QCD calculation of the Isgur-Wise function, and on the corresponding determination of $V_{cb}$. An approach complementary to the one described here is currently being pursued by Mandula and Ogilvie [3] and by Aglietti [4]. These authors are attempting to formulate the heavy quark effective theory in Euclidean space, and to exploit this formulation for a numerical evaluation of the Isgur-Wise function.

To obtain the Isgur-Wise function, we evaluate the elastic scattering matrix element $\langle D(p')|\bar{c}\gamma^\mu c|D(p)\rangle$ on the mass shell [5]. Because the electromagnetic current $\bar{c}\gamma^\mu c$ is conserved, this matrix element can be parametrized in terms of a single form factor:

$$\langle D(p')|\bar{c}\gamma^\mu c|D(p)\rangle = m_D (v+v')^\mu \, h^{\text{el}}(w) \, , \qquad (1)$$

where $p^{(\prime)} = m_D v^{(\prime)}$ and $\omega = v \cdot v'$ is the four-velocity recoil. In the limit of exact heavy-quark symmetry this form factor is simply $\xi(\omega)$.

There are two sources of corrections to this simple result:

$$h^{\text{el}}(\omega) = \left[1 + \beta^{\text{el}}(\omega) + \gamma^{\text{el}}(\omega)\right] \xi(\omega) \, . \qquad (2)$$

The first correction, $\beta^{\text{el}}(\omega)$, results from radiative corrections to the heavy-quark current. The second correction, $\gamma^{\text{el}}(\omega)$, is due to higher-dimension operators with coefficients proportional to inverse powers of the charm quark mass. As defined in Eq. (2), $\xi(\omega)$ is renormalization-group invariant and normalized to one at $\omega = 1$ [6].

The radiative corrections can be evaluated analytically in a model-independent way: they are perturbative QCD corrections. To obtain these corrections we use Neubert's short distance expansion of heavy-quark currents [6]. Neubert's result accounts for the full order $\alpha_s$ dependence of the heavy-quark current on the mass ratio, $z$, of the current's two heavy quarks. This is important because, in $D \to D$ transitions, $z = 1$ and order $\alpha_s z^n$ corrections, $n = 1, \ldots, \infty$, can be expected to be in the 10% range.

Because $\Lambda_{QCD}/(2m_c) \simeq 1/12$, we would naively expect the corrections which are proportional to inverse powers of the charm quark mass to be of the order of 8%. These corrections are difficult to quantify because they involve the light degrees of freedom and are therefore non-perturbative. Luke's theorem [7], however, guarantees that there are no $\mathcal{O}(\Lambda_{QCD}/(2m_c))$ corrections to $h^{el}(\omega)$ at zero recoil. Moreover, model estimates of these corrections appear to remain well below 3% over the range of experimentally accessible recoils [8]. We will neglect the $\mathcal{O}(\Lambda_{QCD}/(2m_c))$ corrections in extracting the Isgur-Wise function from $h^{el}(\omega)$ and will study the heavy-quark-mass dependence of $h^{el}(\omega)$ in a later publication.

Following Neubert [9], we extract $V_{cb}$ from the experimentally measured differential decay rate for $\bar{B} \to D^* l \bar{\nu}$ decays. Using heavy-quark symmetry, we can express this differential decay rate in terms of $\xi(\omega)$. In the limit of zero lepton mass,

$$\frac{1}{\sqrt{\omega^2 - 1}} \frac{d\Gamma}{d\omega} = \frac{G_F^2}{48\pi^3} m_{D^*}^3 (m_B - m_{D^*})^2 \left[1 + \beta^{A_1}(1)\right]^2 (\omega + 1)^2 |V_{cb}|^2 \xi^2(\omega)$$
$$\times \left[1 + 4\left(\frac{\omega}{\omega + 1}\right) \frac{m_B^2 - 2\omega m_B m_{D^*} + m_{D^*}^2}{(m_B - m_{D^*})^2}\right] K(\omega) , \qquad (3)$$

where $\beta^{A_1}(\omega)$ is the radiative correction corresponding to the form factor relevant for $\bar{B} \to D^*$ transitions at zero recoil; $\beta^{A_1}(1) = -0.01$ [6]. Moreover, up to corrections of order $(\Lambda_{QCD}/(2m_{c,b}))^2$, $K(\omega) = 1$ at zero recoil (Luke's theorem [7]). Away from zero recoil, $K(\omega)$ contains $1/m_{c,b}$ and radiative corrections. Due to a fortunate cancellation, their sum remains small for all values of $\omega$ accessible in semi-leptonic decays [10]. Therefore we will neglect all non-perturbative effects in the coefficients which relate $\xi(\omega)$ to the differential decay rate of Eq. (3), and set $K(\omega) = 1$ for all $\omega$. Then our lattice determination of $\xi(\omega)$ enables us to extract $V_{cb}$ from the experimentally measured differential decay rate.

We work in the quenched approximation on a $24^3 \times 48$ lattice at $\beta = 6.2$, which corresponds to an inverse lattice spacing $a^{-1} = 2.73(5)\,\text{GeV}$, as determined from the string tension [11]. Our calculation is performed on sixty $SU(3)$ gauge field configurations (for details see Ref. [11]). The mesons are composed of a propagating heavy quark with a mass approximately equal to that of the charm quark, and light antiquarks with masses around that of the strange quark. To reduce discretization errors, the quark propagators are calculated using an $\mathcal{O}(a)$-improved action [12]. This improvement is particularly important here since we are studying the propagation of quarks whose bare masses are around one third the inverse lattice spacing. Our statistical errors are calculated according to the bootstrap procedure described in Ref. [11].

The calculation of the matrix element $\langle D(\mathbf{p}')|\bar{c}\gamma^\mu c|D(\mathbf{p})\rangle$ proceeds along lines which are now standard in the field of lattice computations of weak matrix elements. (For a recent review on this subject and references to the original literature see, for example, Ref. [13]). Thus, we calculate the ratio of three-point correlators,

$$A^\mu(t;\mathbf{p}',\mathbf{q}) \equiv \frac{\sum_{\mathbf{x},\mathbf{y}} e^{-i\mathbf{q}\cdot\mathbf{x}} e^{-i\mathbf{p}'\cdot\mathbf{y}} \langle J_D(t_f,\mathbf{y})\, V^\mu(t,\mathbf{x})\, J_D^\dagger(0,\mathbf{0})\rangle}{\sum_{\mathbf{x},\mathbf{y}} \langle J_D(t_f,\mathbf{y})\, V^0(t,\mathbf{x})\, J_D^\dagger(0,\mathbf{0})\rangle}, \qquad (4)$$

where $J_D$ is a spatially-extended interpolating field for the $D$ meson [14] and $V^\mu$ is the $\mathcal{O}(a)$-improved version of the vector current $\bar{c}\gamma^\mu c$ [15]. To evaluate these correlators, we use the standard source method [16]. We choose $t_f = 24$ and symmetrize the correlators about that point using Euclidean time reversal [17]. We evaluate $A^\mu$ for three values of the light-quark mass ($\kappa_l = 0.14144, 0.14226, 0.14262$) which straddle the strange quark mass (given by $\kappa_s = 0.1419(1)$ [18]); one value of the heavy-quark mass approximating that of the charm quark ($\kappa_c = 0.129$ [19]); two values of the final $D$-meson momentum, $\mathbf{p}'$, and six values of the momentum, $\mathbf{p} = \mathbf{q} + \mathbf{p}'$, carried by the initial $D$-meson (these momenta are given in Table I). Data with momenta greater than $(\pi/12a)\sqrt{2}$ are excluded because they have larger statistical and systematic uncertainties. To improve statistics we average over all equivalent momenta.

Provided the three points in the correlators of Eq. (4) are sufficiently separated in time, the ground state contribution to the ratio dominates:

$$A^\mu(t;\mathbf{p}',\mathbf{q}) \xrightarrow[t,t_f-t\to\infty]{} \frac{m_D}{2E_D E'_D} \frac{Z_D(\mathbf{p}^2) Z_D(\mathbf{p}'^2)}{Z_D^2(0)} e^{-(E_D-E'_D)t-(E'_D-m_D)t_f} \langle D(\mathbf{p}')|\bar{c}\gamma^\mu c(0)|D(\mathbf{p})\rangle,$$
(5)

where we have used the fact that, for the continuum current, $\langle D(\mathbf{0})|\bar{c}\gamma^0 c(0)|D(\mathbf{0})\rangle = 2m_D$. $E_D$ ($E'_D$) is the energy of the initial (final) $D$ meson and the wavefunction factor, $Z_D(\mathbf{p}^2) \equiv \langle 0|J_D(0)|D(\mathbf{p})\rangle$, is a function of the meson's momentum, because we use spatially-extended interpolating operators.

We fit to

$$R(t;\mathbf{p}',\mathbf{q}) = A^0(t;\mathbf{p}',\mathbf{q}) \times e^{(E_D-E'_D)t+(E'_D-m_D)t_f}.$$
(6)

This ratio becomes independent of $t$ when both $t$ and $t_f - t$ are sufficiently large, since the exponential factor in Eq. (6) explicitly cancels $A^0$'s time dependence. We see a plateau in $R$ about $t = 12$, and fit $R(t;\mathbf{p}',\mathbf{q})$ to a constant for $t = 11, 12, 13$. Multiplying this constant by suitable wavefunction and energy factors, we obtain $\langle D(\mathbf{p}')|\bar{c}\gamma^0 c|D(\mathbf{p})\rangle$. All wavefunction factors and energies are obtained from fits to two-point functions. Our results for $\xi(\omega)$ are presented in Table I.

The data for $\kappa_l = 0.14144$, the heaviest of our light quarks, which have the smallest statistical errors, are shown in Fig. 1. The solid curve is a two-parameter fit to $s\xi_\rho(\omega)$, where $\xi_\rho(\omega)$ is Stech's relativistic-oscillator parametrization [8,10]:

$$\xi_\rho(\omega) = \frac{2}{\omega+1} \exp\left(-(2\rho^2-1)\frac{\omega-1}{\omega+1}\right)$$
(7)

and $\rho^2 = -\xi'_\rho(1)$. The parameter $s$ was added to absorb uncertainties in the overall normalization of our data through a common factor. We find $\rho^2 = 1.5^{+2}_{-2}$ and $s = 0.95^{+1}_{-1}$ with a $\chi^2/\mathrm{dof} = 1.0$. Other parametrizations for $\xi(\omega)$ give nearly identical results; for instance, the pole ansatz of Ref. [10] yields $\rho^2 = 1.4^{+2}_{-2}$ and $s = 0.95^{+1}_{-1}$ with a $\chi^2/\mathrm{dof} = 1.0$. The fact that $s$ is not quite consistent with 1 indicates that there may be some small systematic uncertainty associated with our choice of normalization, or that the standard parametrizations for $\xi(\omega)$ are not optimal.

Now, if we fit our data to $\xi_\rho(\omega)$ instead of $s\xi_\rho(\omega)$ (an equally valid procedure, in principle, for determining $\rho^2$), we find $\rho^2 = 2.0^{+1}_{-1}$ with a $\chi^2/\mathrm{d}of = 2.6$ (dotted curve in Fig. 1). To accommodate the spread in values for $\rho^2$ given by our two procedures, we assign errors to $\rho^2$ which encompass all values consistent with both procedures. These errors include systematic uncertainties, but only to the extent that the deviation of $s$ from 1, in our first fit, is an indication of systematic errors. The central value we choose for $\rho^2$ is the one given by our first fit since this fit is designed to absorb possible uncertainties in the overall normalization of our data. Thus, for $\kappa_l = 0.14144$ we quote $\rho^2 = 1.5^{+6}_{-2}$.

In Fig. 2 we plot the results obtained from a covariant and linear extrapolation of our data for the three values of the light-quark mass to $\kappa_l = \kappa_{crit}$. These results correspond to a meson composed of a charm quark and a massless antiquark. Since $\xi(\omega)$ is not thought to depend very strongly on the light-quark mass [20], we expect the slope parameter, $\rho^2$, for this data to be close to that found when $\kappa_l = 0.14144$, i.e. when the light antiquark is slightly heavier than the strange quark. This is indeed what we find. With a two parameter fit to $s\xi_\rho(\omega)$, we get $\rho^2 = 1.2^{+3}_{-3}$ and $s = 0.94^{+2}_{-2}$ with a $\chi^2/\mathrm{d}of = 0.9$. Forcing $s$ to be 1, we find $\rho^2 = 1.7^{+2}_{-2}$ with a $\chi^2/\mathrm{d}of = 1.6$. Using the same procedure as the one used above to determine the errors and central value for $\rho^2$ when $\kappa_l = 0.14144$, we obtain, as our best estimate for $\rho^2$, when $\kappa_l = \kappa_{crit}$:

$$\rho^2 = 1.2^{+7}_{-3}. \tag{8}$$

Our result for $\rho^2$ agrees with most other determinations of this parameter [9,21,22], apart from the sum-rule result of Ref. [23] which lies below our error bars. In particular, our result for $\rho^2$ agrees with the very recent lattice result of Ref. [24] obtained with Wilson fermions, although the details and systematics of the two calculations are different.

Having chosen a parametrization for the Isgur-Wise function (Eq. (7)), and having determined the slope parameter (Eq. (8)), we can now obtain $V_{cb}$. Setting $K(\omega)$ to one, as discussed after Eq. (3), we fit the decay rate of Eq. (3) to the experimental data. In Fig. 3 we show a least-$\chi^2$-fit to the new ARGUS data for $\bar{B} \to D^* l \bar{\nu}$ decays [25]. The resulting

value of $V_{cb}$ is

$$|V_{cb}|\sqrt{\frac{\tau_B}{1.48\text{ps}}} = 0.038^{+2+8}_{-2-3}, \qquad (9)$$

with a $\chi^2/\mathrm{d}of = 1.1$. The same fit to the weighted average of older CLEO and ARGUS data [26] gives $|V_{cb}|\sqrt{\tau_B/1.48\text{ps}} = 0.036^{+2+8}_{-2-3}$ with a $\chi^2/\mathrm{d}of = 0.6$. In both cases, the first set of errors is due to experimental uncertainties, while the second is due to the uncertainty in our lattice determination of $\rho^2$. The B-meson lifetime used above is the central value of the lifetime, $\tau_{B^0} = 1.48(10)\text{ps}$, quoted in Ref. [27].

In this letter we have reported on a lattice computation of the Isgur-Wise function and on the corresponding determination of the CKM-matrix element, $V_{cb}$. It should be stressed that the lattice errors for $\rho^2$ in Eq. (8) and for $V_{cb}$ in Eq. (9) include systematic errors only to the extent that the spread in values for $\rho^2$ given by the two procedures discussed after Eq. (7) is a measure of these errors. We have tried to minimize systematics by working with an improved action to reduce discretisation errors, and on a fairly large volume in the hope that finite size effects would be small. Nevertheless, it is important that our simulation be repeated on lattices with different lattice spacings and on lattices of different sizes in order to quantify more precisely these systematic effects. It should also be remembered that the computation was performed in the quenched approximation.

Lattice simulations provide the opportunity for testing the validity of the Heavy Quark Effective Theory for charm quarks, by allowing one to check whether $\xi(\omega)$ is independent of the mass of the heavy quark. They further enable one to check the extent to which the Isgur-Wise function obtained from pseudoscalar $\to$ vector transitions agrees with that obtained from pseudoscalar $\to$ pseudoscalar transitions. Both these checks are currently under investigation.

FIGURES

FIG. 1. The squares are our lattice measurements for the Isgur-Wise function, $\xi(\omega)$, at the heaviest light-quark mass ($\kappa_l = 0.14144$). The solid curve is obtained by fitting our measurements to $s\xi_\rho(\omega)$, where $\xi_\rho(\omega)$ is Stech's relativistic-oscillator parametrization for the Isgur-Wise function (Eq. (7)). The dotted curve is obtained by fitting our data to $\xi_\rho(\omega)$.

FIG. 2. The octogons are our lattice results for the chirally-extrapolated Isgur-Wise function, i.e. $\xi(\omega)$ for $\kappa_l = \kappa_{crit}$. The solid curve is obtained by fitting these results to $s\xi_\rho(\omega)$, where $\xi_\rho(\omega)$ is Stech's relativistic-oscillator parametrization for the Isgur-Wise function (Eq. (7)). The dotted curve is obtained by fitting our results to $\xi_\rho(\omega)$.

FIG. 3. The best fit of $|V_{cb}|\xi_\rho(\omega)$ to experimental data, where $\rho^2$ is fixed to its value in Eq. (8). The experimental data (diamonds) are obtained from ARGUS's new results for $\bar{B} \to D^* l \bar{\nu}$ decays, assuming a $B$ meson lifetime of 1.48 ps [27]. Also shown are our appropriately scaled, chirally-extrapolated results (octogons).

TABLES

TABLE I. Results for the Isgur-Wise function, $\xi(\omega)$. Only data for our heaviest light quark ($\kappa_l = 0.14144$) and for a massless light quark ($\kappa_l = \kappa_{crit} = 0.14315(2)$ [18]) are displayed. The hopping parameter of the heavy quark is $\kappa_h = 0.129$. **p** (**p**′) is the momentum of the initial (final) pseudoscalar meson in lattice units.

| | **p′ = (0,0,0)** | | | |
|---|---|---|---|---|
| $\kappa_l$ | 0.14144 | | $\kappa_{crit}$ | |
| **p** | $\omega$ | $\xi(\omega)$ | $\omega$ | $\xi(\omega)$ |
| (1,0,0) | $1.060^{+2}_{-2}$ | $0.89^{+1}_{-1}$ | $1.070^{+4}_{-4}$ | $0.89^{+2}_{-2}$ |
| (1,1,0) | $1.107^{+6}_{-6}$ | $0.80^{+2}_{-2}$ | $1.117^{+11}_{-11}$ | $0.80^{+4}_{-4}$ |
| | **p′ = (1,0,0)** | | | |
| $\kappa_l$ | 0.14144 | | $\kappa_{crit}$ | |
| **p** | $\omega$ | $\xi(\omega)$ | $\omega$ | $\xi(\omega)$ |
| (1,0,0) | $0.990^{+5}_{-5}$ | $0.94^{+4}_{-4}$ | $0.990^{+8}_{-8}$ | $0.92^{+9}_{-9}$ |
| (1,1,0) | $1.039^{+9}_{-9}$ | $0.88^{+4}_{-4}$ | $1.040^{+14}_{-14}$ | $0.94^{+10}_{-9}$ |
| (0,0,0) | $1.060^{+2}_{-2}$ | $0.86^{+1}_{-1}$ | $1.070^{+4}_{-4}$ | $0.86^{+2}_{-2}$ |
| (0,1,0) | $1.125^{+5}_{-5}$ | $0.79^{+2}_{-2}$ | $1.144^{+8}_{-8}$ | $0.78^{+3}_{-3}$ |
| (0,1,1) | $1.173^{+8}_{-9}$ | $0.72^{+3}_{-3}$ | $1.195^{+14}_{-14}$ | $0.70^{+5}_{-5}$ |
| (−1,0,0) | $1.259^{+4}_{-4}$ | $0.69^{+2}_{-2}$ | $1.298^{+8}_{-8}$ | $0.73^{+4}_{-4}$ |

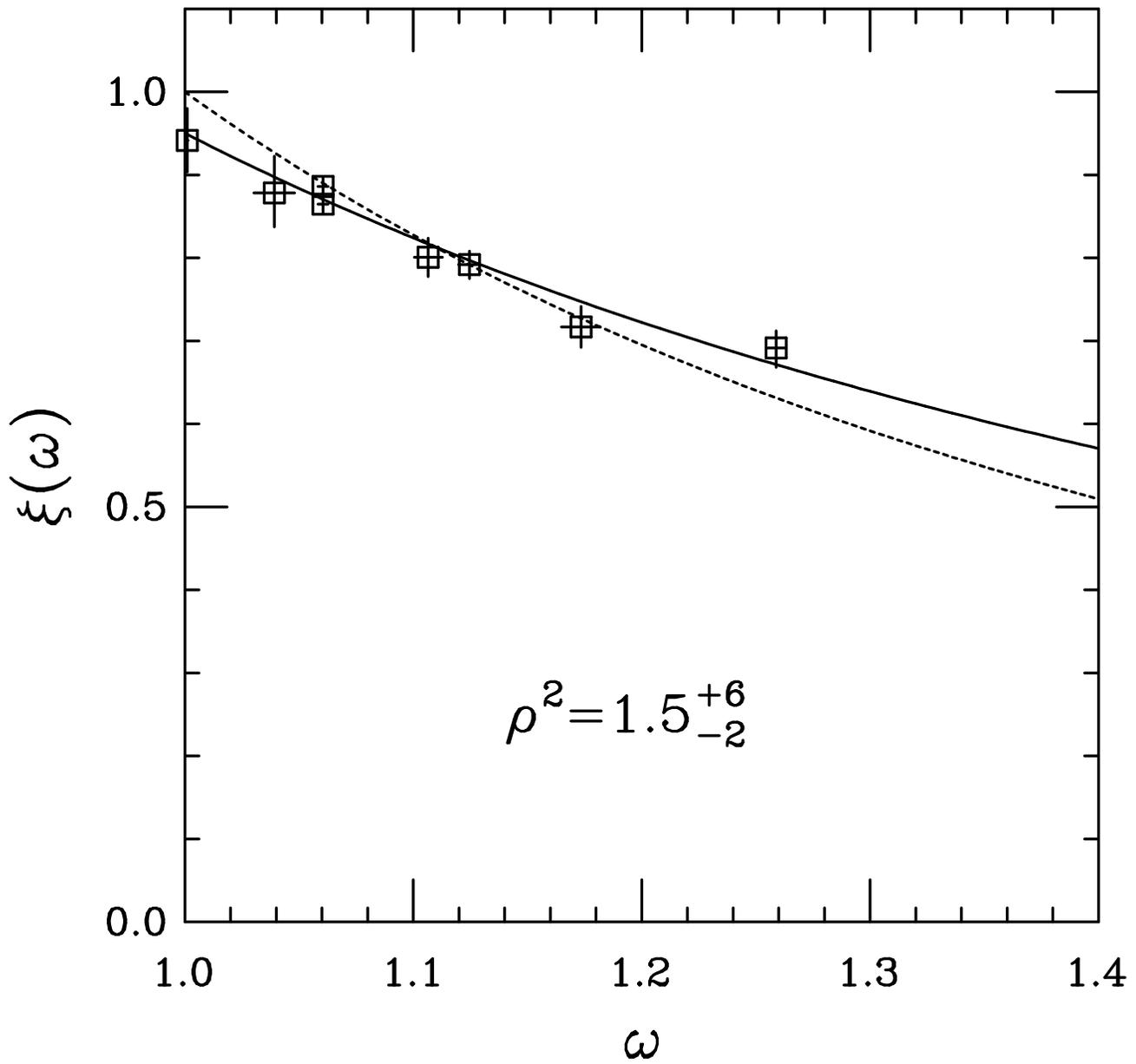

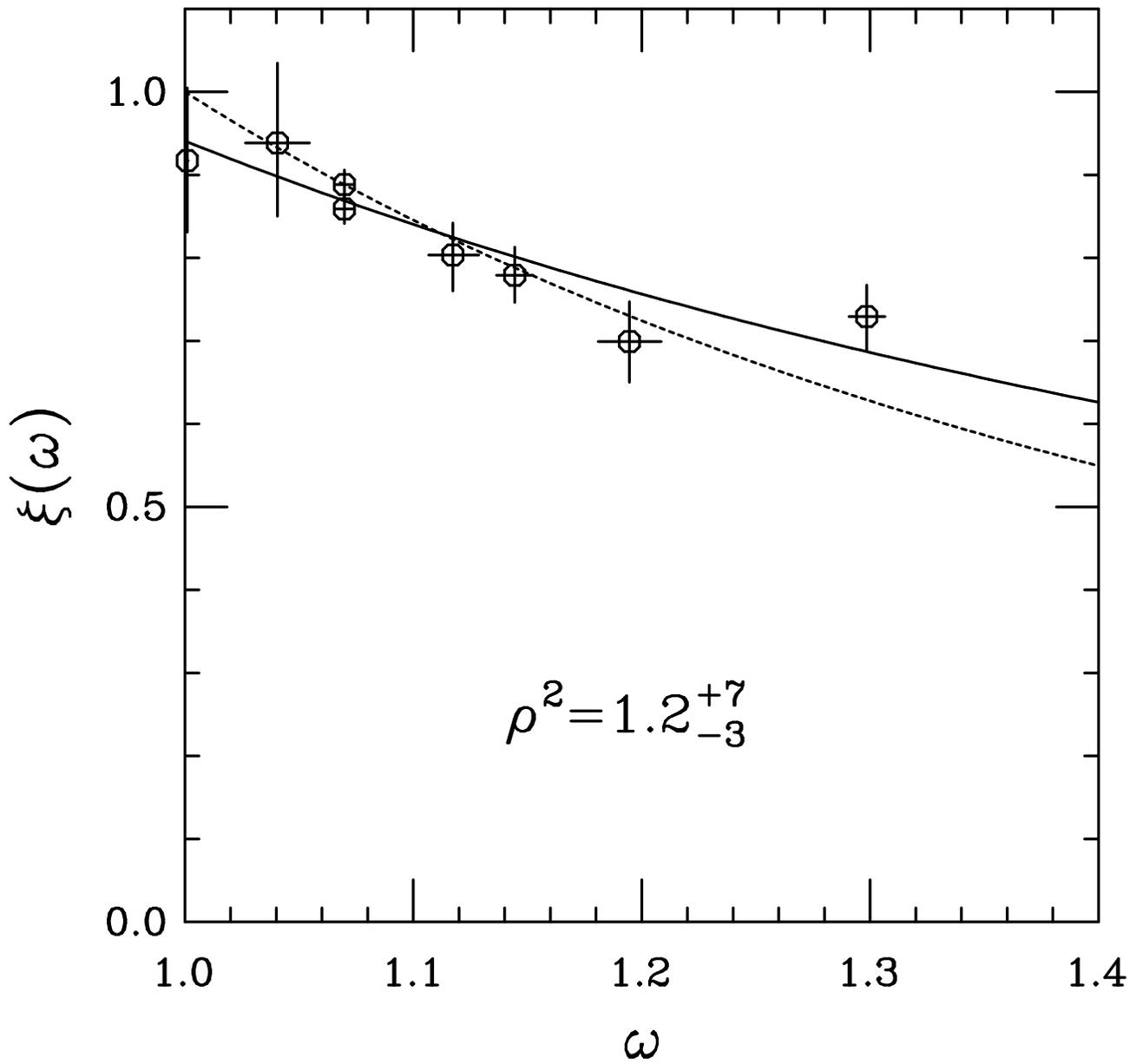

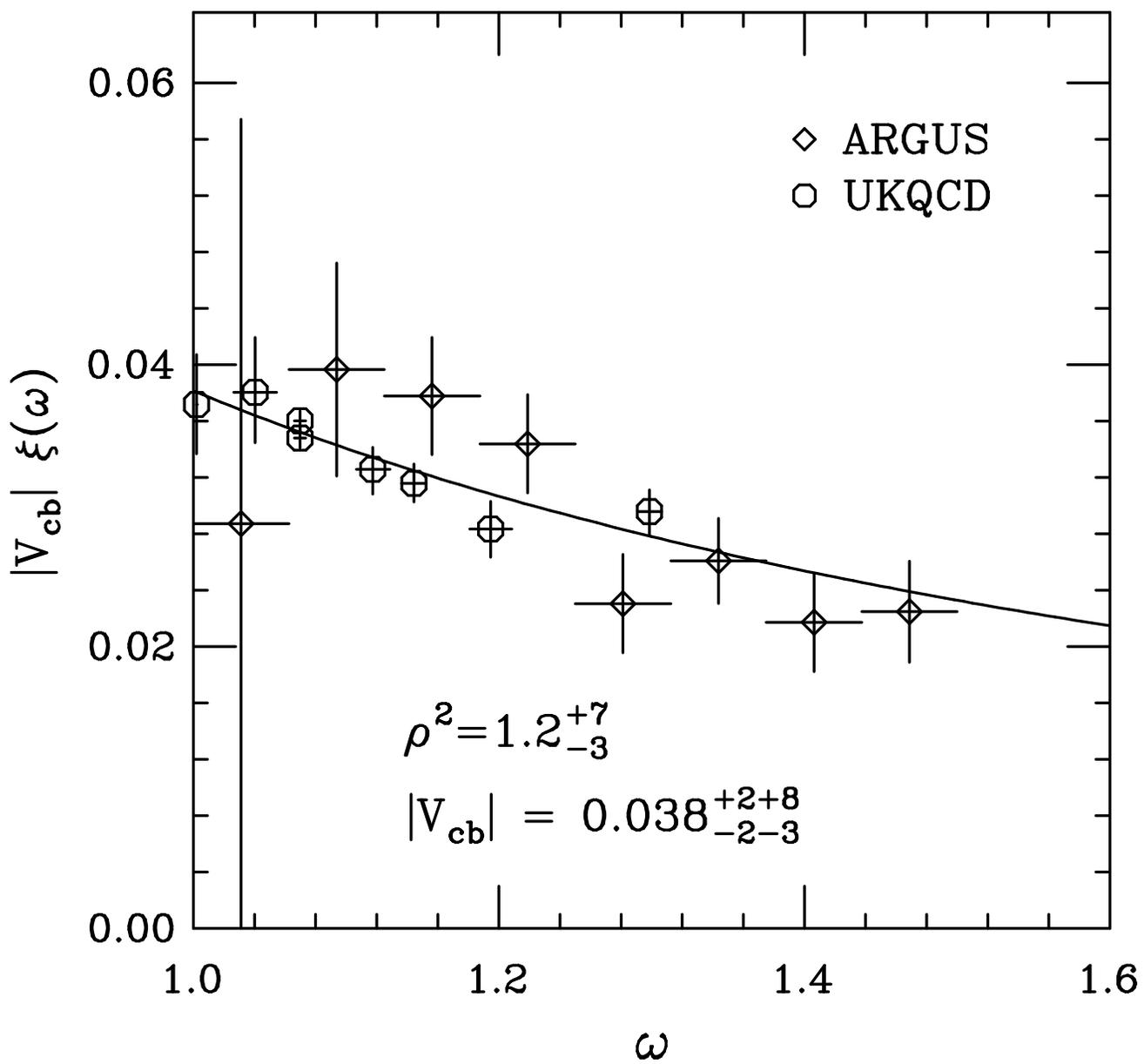